# Aerodynamic performance and flow mechanism of 3D flapping wing using discrete vortex method

**Rahul Kumar** [a], **Srikant S. Padhee** [a] and **Devranjan Samanta** [a,*,1]

[a] *Department of Mechanical Engineering, Indian Institute of Technology Ropar Rupnagar-140001, Punjab, India*

*Corresponding author:

[1]E-mail: devranjan.samanta@iitrpr.ac.in

[1]Tel: +91-1881-24-2109

**Abstract**

In this work, we have performed numerical simulations of the flapping motion of a rectangular wing in a three-dimensional flow field using the discrete vortex method (DVM). The DVM method is computationally more convenient because it does not require the generation of a grid for the flow field at each time step as in other conventional simulation methods. In addition to the rigid wing case, the aerodynamic characteristics of a deformable wing are also investigated. The deformable wing is studied in various configurations, such as bending, twisting, and bending-twisting coupling (BTC), to provide a comprehensive analysis of its performance. In this study, we have introduced a novel aerodynamic technique in wing twisting. Unlike traditional wing rotation about a fixed root axis, our approach involves rotating the wing about a dynamically adjusted point located at the root of the leading edge. This novel approach has been employed to enhance the pressure levels generated by the wing, thereby reflecting an increase in the requisite aerodynamic force. The BTC wing represents an innovative aerodynamic approach that optimally coordinates both twisting and bending deformations of the wing, resulting in a substantial improvement in its overall aerodynamic efficiency. The investigation of all four modes involves a detailed analysis of the flow mechanisms and vortex dynamics, which play a crucial role in influencing the aerodynamic forces, namely lift and thrust. The study aims to understand how these flow patterns change under different operating conditions and how these changes impact the generation of lift and thrust. The lift, thrust, and propulsive efficiency of all four modes are compared to provide a detailed understanding of their aerodynamic characteristics. The bent wing showed more lift compared to the rigid wing, but minimal improvements in thrust. In contrast, the twisted wing showed greater improvements in both lift and thrust. The BTC wing proves to be the most efficient method to improve aerodynamic performance during flapping. The parametric dependence of kinematic parameters such as asymmetric ratio (downstroke speed to upstroke speed), aspect ratio and reduced frequency on the aerodynamic performance was also investigated.

## 1. Introduction

Drones and micro air vehicles are ubiquitous for aerial surveillance in modern-day warfare and natural disaster management. To develop small flapping drones, biomimicry of flapping flight in insects and birds needs to be studied. The advancement of a micro air vehicles (MAVs) has greatly benefited from the numerous computational and experimental researches that have been performed in recent years to examine the aerodynamic performance of insects [1]–[6]. As the length scale of the flying body decreases, flapping mechanisms seem to be more efficient than fixed-wing aerodynamics. The flapping flight mechanism has been an area of interest since the pioneering work of Lighthill [7] and Ellington [8]. Early investigations on bird and insect flight have been well-covered in various books and review papers [9], [10]. The smallest UAVs, also known as nano air vehicles, typically have centimetre-scale dimensions and may weigh less than 100 mg [11], [12]. They fly at Reynolds numbers approximately below 1000, where the aerodynamic performance of traditional lifting surfaces (like fixed and rotating wings) diminishes due to significant viscous effects in the flow that encourage flow detachment and friction. Therefore, the idea of utilizing bio-inspired flapping wings has generated curiosity as a possible substitute for conventional fixed and rotating wings on UAVs of incredibly small dimensions. This interest stems from the fact that certain insects operate within a comparable range of Reynolds numbers. Furthermore, in contrast to fixed wings, flapping wings provide hovering flight and may offer further advantages in terms of manoeuvrability and noise emission [5]. Unsteady flapping dynamics has various features, such as clap and fling mechanism, delayed stall, and wake capture during collective motion [13]. Recently, chaotic dynamics of the wake behind flapping wings were reported. This study explained how the complex interactions between the primary vortex structures contribute to the unsteady flow-transition field from periodicity to chaos [14]. The forward thrust generation mechanism due to flapping was elucidated by a numerical simulation study [15]. Numerical simulations of a 1-D elastic solid in a 2-D flow were done to reveal the interplay of vorticity and elastodynamics in flapping flight [16].

In comparison to the study of three-dimensional rigid flapping wings, the role of wing deformation in improving aerodynamic performance is a relatively lesser-studied problem [17]–[19] . Insect wings that flap are extremely flexible structures that bend when subjected to aerodynamic and inertial stresses [20]. Deformation has been demonstrated to improve aerodynamic force generation [21]–[23] and efficiency [24]–[26], and it is thought to play an important role in the mechanics of insect flying. Venation and corrugation, among other structural characteristics, have an impact on how wings deform when subjected to dynamic loads[27]. These characteristics also control the spatial variation of the mechanical characteristics of the wing. According to experimental research, the flexural rigidity of the forewing of the hawkmoth Manduca sexta, for instance, changes by one to two orders of magnitude along the chord length of the wing [28]. This strong gradient may support wing responses that are advantageous from an aerodynamic and energy standpoint. However, it is unclear how graded flexural stiffness affects the performance of flapping wings. Yin and Luo [26] utilized a pitch-plunge model to demonstrate that a wing's power efficiency was higher when viscous fluxes, as opposed to inertia, were the dominant cause of deformation. Using this conceptual framework, The impact of the wing's deformation during the upstroke and downstroke on forward flight was studied by Tian *et al.* [29], who observed that flexible wings generate a higher lift range at the cost of increased inertial power inputs. At low Reynolds numbers, Vanella *et al.* [21] showed that a super-harmonic resonance may positively affect the wing's lift-to-drag ratio. Mountcastle and Daniel [22] looked into the effects of varying wing kinematics and structural characteristics on the generation of lift from the wing. Joshi and Mysa [30] highlighted the importance of performance enhancement in tandem flapping wings, finding out that the hind foil in this configuration can even generate approximately twice as much thrust as the fore/single flapping foil.

Unsteady separated flows have also been extensively modelled using discrete-vortex method (DVM). These techniques are primarily based on potential-flow theory, and the discrete vortices that emerge from the surface indicate the shear layers that represent the separated flow. Saffman & Baker [31] and Clements & Maull [32] provide thorough background information on the development of the discrete-vortex approach over time. Leonard [33] provides a discussion of more recent developments regarding the use of vortex algorithms for flow simulation. This method has been successfully used to predict flow past inclined plates and bluff bodies [34]–[36]. Katz [37] created a method for partially separated flow past an airfoil, although it requires experimentation or other methods to determine where the separation occurs on the aerofoil. More recently, leading-edge vortices in unsteady flows have been modelled using low-order techniques based on discrete vortices [38]–[41]. While these techniques rely on potential theory, they effectively incorporate the fundamental principles of fluid dynamics in relevant flow scenarios through the integration of discrete-vortex shedding, thereby augmenting inviscid theory. High-speed computation and experimental equipment advancements have made it possible to take into account the effects of additional variables on the aerodynamic efficiency of flapping wings, including the three-dimensional effect [42], [43], hovering [43], [44], and interactions between multiple wings [45], [46]. This has made flapping-wing models more detailed and realistic to real flight [47] in the natural world. In our previous work [48]–[50], we simulated the flapping of a one-dimensional (1-D) flexible filament in a two-dimensional (2-D) inviscid flow. The performance of rigid and flexible wings for symmetric and asymmetric flapping was compared. We have also shown that the variation of wing length during a flapping cycle is a useful strategy in augmenting the lift performance.

The present article highlights the 3-D numerical simulation results of flapping wings using DVM. In the 3D DVM technique, only the wing is discretized. Since the flow domain is not discretized, the effort of discretizing the time-varying flow domain is saved for unsteady flapping simulation at each time step. We have performed a comparative study of aerodynamic performance of different configurations like rigid, bent, twisted wings [51]–[53] and finally bending-twisting-coupling [BTC]. The bending and twisting modes are implemented using spanwise and chordwise travelling wave motion. The parametric dependence of lift and thrust for these 4 configurations for different kinematic parameters like aspect ratio ($A_R$), asymmetry ratio ($A_S$=ratio of downstroke speed to upstroke speed) and reduced frequency ($k$) are reported here. The investigation of all four modes involves a thorough analysis of flow mechanics and vortex dynamics, fundamental factors that profoundly influence aerodynamic forces like lift and thrust. The main goal of this study is to identify how flow patterns change under different operating conditions and understand their direct impact on lift and thrust generation. This rigorous analysis includes a detailed examination of vortex structures, such as trailing-edge vortices and other coherent flow features, to understand how they evolve and behave in each mode. By delving into these vortex dynamics, researchers can unveil the underlying physical mechanisms governing lift and thrust generation for each specific mode. This study aims to expand upon prior investigations, including experimental work [10] and numerical simulations [54]–[56], where researchers examined the aerodynamic performance and flow behaviour associated with flapping mechanisms. We believe that our simulation results will provide a framework for designing flapping MAVs with deformable wings.

## 2. Computational model

In this analysis, we present a three-dimensional flapping simulation using the discrete vortex method (DVM). This method is highly accurate in analyzing unsteady potential flow [37], [41], [50], [57]–[59]. In this method, the wing was discretized into numerous quadrilateral flat surface panels (Fig. 1a). So, the wing was discretized into $N \times M$ panels, with $N$ panels along the span and M panels along the chord. The wing's bound circulation and the vortex wake were modelled using vortex ring elements. At each quarter of the panel, leading segment of the vortex ring was located (Fig. 1b). The collocation point, where no normal flow conditions were applied, was located at the centre of three-fourths of the panel. The normal vector, n, was defined at the collocation point, which also served as the centre of the vortex ring for each panel. At any given point ($x, y, z$), three components of velocity ($u, v, w$) can be defined according to the vortex ring formulation (eq-3). Initially, there would be a vortex ring only on the wing panels. The trailing vortex segment of the trailing edge served as the starting vortex during the first-time step, since there were no free wake elements available. The trailing edge of the wing advanced during the second time step, and by combining the new aft points of the trailing edge vortex ring. With each time step, the shedding process produces a new set of trailing edge wake vortex rings (known as the wake shedding procedure). By employing Kelvin's circulation theorem on inviscid flows, at each time step, the trailing edge wake panel sheds a wake panel with vortex strength equal to its circulation from the preceding step. The strength of the wake shed was constant over time because we neglected the vortex diffusion terms in our simulations.

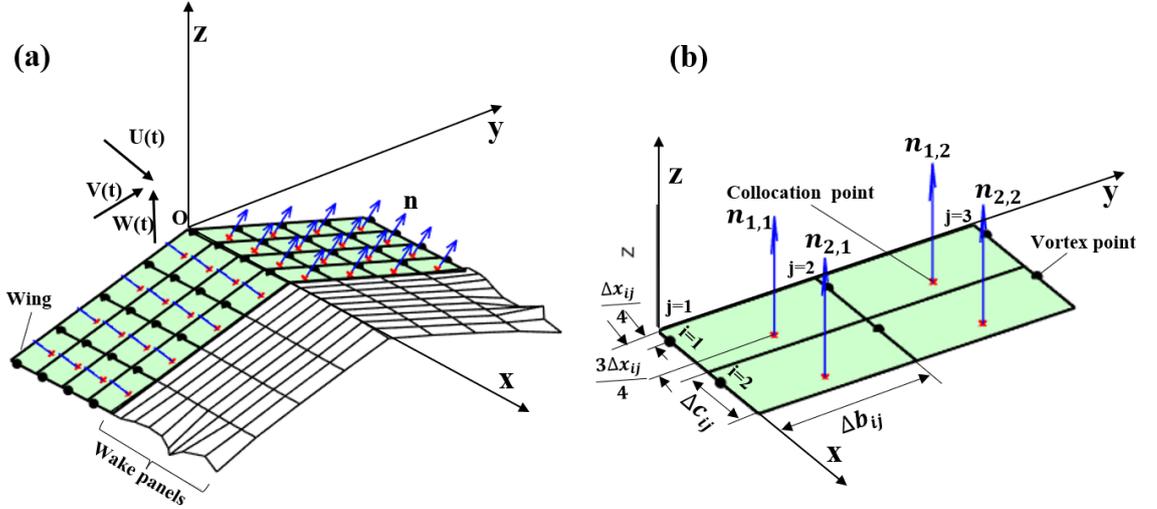

Fig. 1. (a) Illustration of wing and wake panel (b) Illustration of vortex and collocation point. At each quarter of the panel leading segment of the vortex ring is located. The collocation point where no normal flow condition is applied is located at the center of the three fourth of the panel.

### 2.1 Implementation of discrete vortex method

The governing equation is based on the vorticity transport equation:

$$\frac{\partial \Omega}{\partial t} + \vec{u} \cdot \nabla \Omega = \Omega \cdot \nabla \vec{u} + v \nabla^2 \Omega \qquad (1)$$

The velocity field in equation (1) is represented by $\vec{u}$ ($x, t$) and the vorticity field by $\Omega$ (x, t) = $\nabla \times \vec{u}$ (x, t). The vortex diffusion term ($v \nabla^2 \Omega$) is eliminated as a result of the inviscid flow assumption. Contrary to the 2-D flow field [48], [49], vortex stretching and tilting term is an additional complexity in 3-D flow computations. Consequently, the modified equation is as follows:

$$\frac{D\Omega}{Dt} = \Omega \cdot \nabla \vec{u} \qquad (2)$$

$D/Dt$ is a material derivative of vorticity in equation (2). The Lagrangian approach is used in this numerical scheme to track the wakes that are produced due to flapping motion. For point vortices, at very close distances between each other, there is possibility of generation of singularities.

The expression governing the Biot-Savart law for a finite vortex segment is as follows.

$$\vec{dv} = \frac{\Gamma}{4\pi}\left[\frac{\vec{dl} \times \vec{r}}{|\vec{r}|^3}\right] \qquad (3)$$

where, $\vec{dv}$ is the induced velocity caused by a vortex segment with a length $\vec{dl}$. The velocity is determined by the circulation strength $\Gamma$ and the distance r between the segment and the point of interest.

The normal velocity induced at the collocation point by a panel of unit circulation strength is referred to as the influence coefficient $a_{KL}$. Therefore, at the $K^{th}$ collocation point, for a vortex element $L$ of unit strength, the influence coefficient is described as:

$$a_{KL} = (u, v, w)_{KL} \cdot n_K, \qquad (4)$$

where $K$ and $L$ are the vertical and horizontal matrix counters, respectively. The order of the matrix is $m = M \times N$. On any randomly chosen collocation point $K$, we obtain the zero normal velocity boundary condition on the surface ($Q_{nK} = 0$).

$$Q_{nK} = a_{K1}\Gamma_1 + a_{K2}\Gamma_2 + \cdots + a_{Km}\Gamma_m + [U(t) + u_w, V(t) + v_w, W(t) + w_W]_K \cdot n_K = 0, \qquad (5)$$

The equation on the right-hand side deals with the normal flow velocity induced by previously shedded wakes and wing motion:

$$RHS_K = -[U(t) + u_w, V(t) + v_w, W(t) + w_W]_K \cdot n_K, \qquad (6)$$

where $(u, v, w)_w$ is the induced velocity caused at the $K^{th}$ panel by the complete wake and $U, V,$ and $W$ are the time-dependent kinematic velocities of the wing. The aerodynamic loads can be calculated once the new values for wing circulation have been identified. The unsteady Bernoulli equation is utilized to determine the pressure distribution and, consequently, the lift and drag coefficient of the wing.

The normal vector at the $K^{th}$ collocation point is denoted by the expression $(\sin\alpha_K, \cos\alpha_K)$ in Fig. 1. Thus, the set of algebraic equations is arranged as follows:

$$\begin{bmatrix} a_{11} & a_{12} & \cdots & a_{1m} \\ a_{21} & a_{22} & \cdots & a_{2m} \\ \cdot & & & \cdot \\ \cdot & & & \cdot \\ \cdot & & & \cdot \\ a_{m1} & a_{m2} & \cdots & a_{mm} \end{bmatrix} \begin{Bmatrix} \Gamma_1 \\ \Gamma_2 \\ \cdot \\ \cdot \\ \cdot \\ \Gamma_m \end{Bmatrix} = \begin{Bmatrix} RHS_1 \\ RHS_2 \\ \cdot \\ \cdot \\ \cdot \\ RHS_m \end{Bmatrix}, \quad (7)$$

The circulation strength of the rings attached to the surface must be calculated at each time step by solving the matrix (7).

The results of this matrix equation can be summarized in indicial form (for each collocation point $K$) as

$$\sum_{L=1}^{m} a_{KL} \Gamma_L = RHS_K, \quad (8)$$

where the coefficients of the inverted matrix are $a_{KL}$.

The aerodynamic loads can be calculated once the new values for wing circulation have been identified. The unsteady Bernoulli equation is utilized to determine the pressure distribution and, consequently, the lift and drag coefficient of the wing.

$$\frac{p_{ref} - p}{\rho} = \frac{Q^2}{2} - \frac{v_{ref}^2}{2} + \frac{\partial \phi}{\partial t}, \quad (9)$$

where the subscript $ref$ denotes the far-field reference conditions and $Q$ represents the local fluid velocity (including kinematic and induced velocities). The symbol $\phi$ denotes the velocity potential.

## 2.2 Aerodynamic loads

According to Katz's book on DVM methods [58], the pressure difference between the airfoil's upper and lower surfaces can be stated as follows:

$$\Delta p_{ij} = \rho \left\{ [U(t) + u_w, V(t) + v_w, W(t) + w_W]_{ij} \cdot \tau_i \frac{\Gamma_{i,j} - \Gamma_{i-1,j}}{\Delta c_{ij}} \right.$$
$$\left. + [U(t) + u_w, V(t) + v_w, W(t) + w_W]_{ij} \cdot \tau_j \frac{\Gamma_{i,j} - \Gamma_{i,j-1}}{\Delta b_{ij}} + \frac{\partial}{\partial t} \Gamma_{i,j} \right\}, \quad (10)$$

where, $\Delta c$ and $\Delta b$, are the panel chord and span length, respectively, while $\tau_i$ and $\tau_j$, represent the tangential vectors of the panel in the $i$ and $j$ directions. The first two terms, respectively, signify the spanwise and chordwise parts of the tangential velocity caused by the wing vortices. The velocity potential time derivative is represented by the third term. This third term is crucial for effectively modeling flapping flying at high reduced frequencies[58].

The total lift at a panel location ($i,j$) is defined as:

$$L = \sum_{j=1}^{N} \sum_{i=1}^{M} \Delta L_{ij} = \sum_{j=1}^{N} \sum_{i=1}^{M} \Delta p_{ij} S_{ij} \cos \alpha_{ij}, \quad (11)$$

The circulation of the most inboard filaments of the body (the filaments closest to the centerline of the body) is zero. By making this assumption, one can create an implicit uniform lift distribution over the body, meaning that the lift per unit span is the same over the entire body as it is over the wings.

The induced drag of the three-dimensional lifting surface at a panel $(i,j)$ is defined as:

$$D = \sum_{j=1}^{N}\sum_{i=1}^{M} \Delta D_{ij}$$

$$= \sum_{j=1}^{N}\sum_{i=1}^{M} \rho\left[(w_{ind} + w_W)_{ij}(\Gamma_{i,j} - \Gamma_{i-1,j})\Delta b_{ij} + \frac{\partial(\Gamma_{ij})}{\partial t}\Delta S_{ij}\sin\alpha_{ij}\right], \quad (12)$$

where $w_{ind}$ is the wake-induced downwash at the three-quarter of each wing panel, The panel's surface area is $\Delta S$, and the angle of attack is $\alpha$. $\rho$ is the density of air. All of the chordwise segments of the vortex rings attached to the wing contribute to the wind's velocity, which is calculated while determining the matrix of influence coefficients. Summation of all the individual panel lifts (eqn. 11) and drags (eqn. 12) will generate the total lift and drag values. The coefficients of lift and thrust ($C_L$, $C_D = L, D/0.5\rho S U_\infty^2$) are defined as total lift or drag normalized by ($0.5\rho S U_\infty^2$).

When an object moves at a constant speed through still air, and the drag force is the only significant force acting on the object in the horizontal direction. The opposite of the drag force can be regarded as the thrust force. Therefore, the coefficient of thrust $C_T$ is basically the negative value of $C_D$. The pressure coefficient, $C_p = p/(1/2\ \rho U_\infty^2)$ can be obtained by summing up the pressure values at individual panels $(i,j)$ using equation (10). To conduct this study, both the work necessary to morph the wing and the inertial power input necessary to flap the wing are disregarded (a valid assumption for forward flight [60]). The average flapping wing's propelling efficiency [61], [62] is denoted as:

$$\eta_p = \frac{\bar{C}_T}{\bar{C}_p}, \quad (13)$$

Where the $\bar{C}_T$ and $\bar{C}_p$ defines the time averaged lift and pressure force.

## 3. Wing kinematics
### 3.1 Wing motion

A wing's motion over a single cycle can be described by angular oscillations. The flapping motions are based on sinusoidal functions, that is,

For symmetric case: $\quad \delta(t) = \delta_0 + \delta_1 \sin(\omega t), \quad (14)$

For asymmetric case: $\quad \delta(t) = \begin{cases} \delta_0 + \delta_1 \sin(\omega t), & 0 \le t < 2/3\ T_p \\ \delta_0 + \delta_1 \sin(A_S * \omega t), & 2/3\ T_p \le t \le T_p \end{cases} \quad (15)$

where, $\delta_0, \delta_1\ and\ \omega$ represents the mean angle, angular amplitude, and angular frequency of flapping, respectively. $t$ is the time in sec and $T_p$ is the period of a flapping cycle. Asymmetric ratio ($A_S$) is defined as the ratio of the downward stroke velocity and the upward stroke velocity.

The wing motion is described by parameters like frequencies, amplitudes, and phase angles (eqn. 14 and 15). At a frequency of 1 Hz, the flapping period ($T_p$) is 1 sec when the wing is symmetric, while in the asymmetric case ($A_R$=2), the flapping period is 0.75 sec. The impact of unsteady flow on the aerodynamic performance of flapping wings is examined using various reduced frequencies. Reduced frequency is a non-dimensional number relevant to unsteady aerodynamics. The reduced frequency *k* is defined as:
$$k = \omega(c/2)/U_\infty, \quad (16)$$
where $\omega$ and $U_\infty$ are the angular frequency and free stream velocity, respectively. c is the chord i.e., equal to 0.1m [55].

Table 1. Parameters used in the current investigation and their possible ranges of variation.

| k | $A_R$ | $A_S$ | $\delta_1$ | $\delta_0$ | $\alpha$ | $U_\infty$ |
|---|---|---|---|---|---|---|
| 0.1,0.2,0.3, 0.4,0.5,0.6 | 1,2,3,4,5,6 | 1,2,3,4,5 | 30° | -30° | 2° | 11m/s |

where $A_R$ is the aspect ratio defined as the ratio of span length (*b*) to chord length (*c*), *k* is the reduced frequency. To calculate the reduced frequency in the case of asymmetric flapping wings, the upstroke frequency is considered. In this numerical study, we have selected the flapping amplitude ($\delta_1$) equal to 30°, initial inclination ($\delta_0$) to a value of -30°, angle of attack ($\alpha$) to a value of 2° and free stream velocity ($U_\infty$) to a value of 11 m/s. Similar values have been adopted in several past studies [5], [63]–[66]. Each time step ($\Delta t$) is taken as 0.001 sec. At each time step, a wake vortex is shed at a distance of $0.3U_\infty\Delta t$ from the wing tip[58].

There are four parts to the wing displacement during forward flight: flapping, $\delta(t)$ (rotation about the wing root), twisting $\theta(X,t)$, bending $def(Y,t)$ and coupling of twisting and bending $(\theta(X,t) + def(Y,t))$ movements. Here, we observe that bending and twisting is a local motion but flapping is a global action (i.e., the entire wing moves as a whole). We introduce two coordinate systems to express wing kinematics. The body frame is the deformed geometry, and it is described by a moving coordinate system called $x_2y_2z_2$. The ground reference frame is described by a fixed coordinate system called $xyz$.

The *x* and *y* axis represent the chordwise and spanwise direction respectively. The positive z-axis is perpendicular to the lifting surface and pointing upwards. Unit vectors along the *x, y,* and *z* axes are denoted by $i_x$, $i_y$, and $i_z$, respectively. Unit vectors along the $x_2$, $y_2$, and $z_2$ axes are denoted by $i_{x_2}, i_{y_2},$ and $i_{z_2}$, respectively. To represent the motion of the wing, we use the two consecutive angles $\delta$ (flapping angle) and $\alpha$ (geometric angle of attack).

Performing a sequence of rotations around the axes in the specified order results in the establishment of the body frame denoted as $x_2y_2z_2$. A pitch-like rotation by $\alpha$ is first performed on the *xyz* system to create the intermediate coordinate system $x_2y_2z_2$ ($y = y_1$). These two coordinate systems' unit vectors are connected by:

$$\begin{pmatrix} i_{x_1} \\ i_{y_1} \\ i_{z_1} \end{pmatrix} = \begin{pmatrix} \cos(\alpha) & 0 & \sin(\alpha) \\ 0 & 1 & 0 \\ -\sin(\alpha) & 0 & \cos(\alpha) \end{pmatrix} \begin{pmatrix} i_x \\ i_y \\ i_z \end{pmatrix}, \quad (17)$$

Second, a flap-like motion is applied to the $x_1y_1z_1$ system to create the second intermediate coordinate system, $x_2y_2z_2$ ($x_1 = x_2$). These two coordinate systems' unit vectors are connected by:

$$\begin{pmatrix} i_{x_2} \\ i_{y_2} \\ i_{z_2} \end{pmatrix} = \begin{pmatrix} 1 & 0 & 0 \\ 0 & \cos(\delta) & \sin(\delta) \\ 0 & -\sin(\delta) & \cos(\delta) \end{pmatrix} \begin{pmatrix} i_{x_1} \\ i_{y_1} \\ i_{z_1} \end{pmatrix}, \quad (18)$$

Consequently, the relationship between the unit vectors of the two coordinate systems, $xyz$ and $x_2 y_2 z_2$, is:

$$\begin{pmatrix} i_{x_2} \\ i_{y_2} \\ i_{x_2} \end{pmatrix} = f(\alpha, \delta) \begin{pmatrix} i_x \\ i_y \\ i_z \end{pmatrix}, \quad (19)$$

The conversion between two coordinate systems is illustrated by the above equation, where the conversion is a function of $\alpha$ and $\delta$.

### 3.2 Wing deformation (active morphing of wing shape)

Fig.2a, b and c depict the shape of bending, twisting and BTC modes respectively. The objective behind implementing this dynamic shape alteration is to explore and examine how the different configurations of deformable wings can improve the wing's flight performance. At the initiation of the downstroke, the wing maintains a flat configuration, gradually transitioning into sinusoidal deformation during the downward stroke, and subsequently reverting to its flat state sinusoidally during the upstroke. The goal of wing twist is to mitigate negative consequences, such as flow separation [64], at high effective angles of attack caused by flap velocity. In the context of this study, we have introduced an innovative aerodynamic technique for wing twisting. Diverging from the conventional practice of wing rotation about root[61], our approach entails the dynamic rotation of the wing about a pivot point (O) situated precisely at the root of the leading edge (Fig.2b). During the stroke, the wing is twisted upward, creating a significant positive angle of attack. Along the span length, the twisting angle ($\theta$) must also change. As a result, maintaining lift in flight requires an effective angle of attack that varies along the wing span (and in maintaining attached flow). The adoption of twisting mode during flapping motion is also observed in a variety of flying species [67]. The BTC wing is characterized by the synergistic integration of both wing twisting and wing bending features. This unique combination of deformations allows for a highly coordinated and optimized performance of the wing. Twisting and bending work together to enhance the wing's aerodynamic efficiency, enabling it to operate more effectively in various flight conditions. The wing undergoes a continuous transformation in its shape, beginning with a flat configuration and progressively increasing its deformation during the upstroke until it reaches maximum deformation at the end of the upstroke. Subsequently, the deformation decreases during the downstroke.

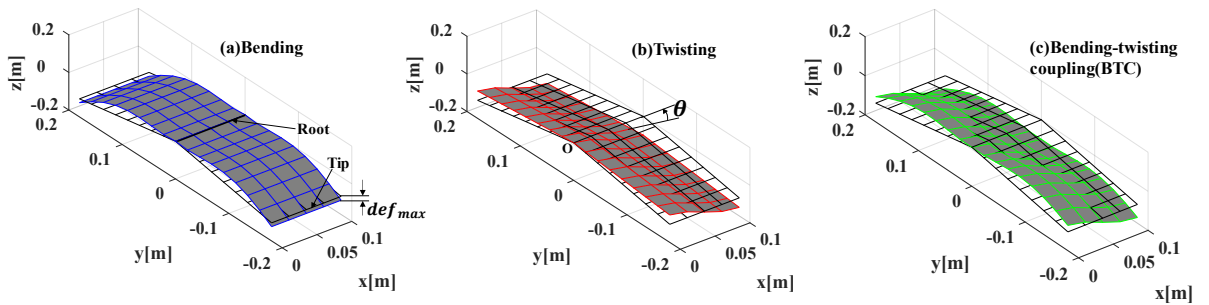

Fig. 2. Different wing shape modes are considered in this work (a) bending (b) twisting (c) bending-twisting coupling (BTC).

Initially, we adopted the approach of Stanford and Beran [61] and consider a predefined deformation of the wing using bending mode shapes. Bending mode is defined as:

$$i_{zb}(t) = i_{z_2} - def_{max}(t) * \left(-192\left(\frac{i_{y_2}}{b}\right)^4 + 224\left(\frac{i_{y_2}}{b}\right)^3 - 60\left(\frac{i_y}{b}\right)^2\right), \quad (20)$$

In the twisting case, we follow the same equation as bending. However, contrary to the previous study by Stanford and Beran [61], the whole wing is twisted about a point "o", as shown in Fig. 2b.
Twisting mode is defined as

$$i_{zt}(t) = -i_{z_2} + \cos(\theta_{max})(t) * \left(-192\left(\frac{i_{x_2}}{b}\right)^4 + 224\left(\frac{i_{x_2}}{b}\right)^3 - 60\left(\frac{i_{x_2}}{b}\right)^2\right), \quad (21)$$

Where $(def_{max}(t), \cos(\theta_{max})(t)) = \begin{cases} (m_0, n_0)b\sin(\omega t), & 0 \leq t < 1/2\, T_p \\ -(m_0, n_0)b\sin(\omega t), & 1/2\, T_p \leq t \leq T_p \end{cases}$

The wing's deformation, denoted by $m_0$ and $n_0$, can be determined based on the specific bending and twisting mode being considered. For the bending mode, $m_0$ is equal to the maximum deflection ($def_{max}$) divided by the wing span ($b$), where the maximum deflection is equal to $0.3b$. On the other hand, for the twisting mode, $n_0$ is equal to the maximum twist ($\cos(\theta_{max})$) divided by wing span ($b$) which has a value of 20°. These values have been adopted from the earlier study [61].

## 4. Validation and results

This section presents the validation of the adopted numerical scheme with Euler's method [51] and unsteady vortex lattice method [61]. The aerodynamic performance of a rigid wing with the addition of bending and twisting modes was numerically investigated to estimate the lift and thrust forces. The wing shape is established as a rectangular planform with a chord length of 0.1 m, as earlier reported [55]. Over the chord, eight panels are employed, and along the span, seventeen panels are used (root to tip). We have performed a parametric dependence study of several kinematic factors, such as reduced frequencies ($k$), aspect ratios ($A_R$), and asymmetric ratios ($A_S$) on the lift and thrust coefficients of the wing. The asymmetry ratio ($A_S$) is the ratio of upstroke speed to downstroke speed. In the case of symmetric flapping, since both upstroke and downstroke speeds are of the same magnitude, the asymmetry ratio $A_S$ is 1. In the case of asymmetric flapping, the downstroke speed is higher than the upstroke speed.

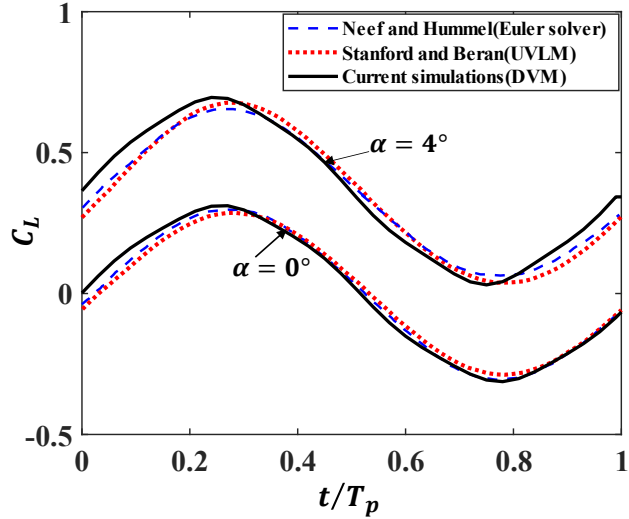

Fig. 3. Comparison of lift for a flapping wing with results from the Euler method [51] and unsteady vortex lattice method(UVLM) [61].

We examined a flapping wing previously studied by Neef and Hummel [51] and studied by Stanford and Beran [61] to demonstrate the DVM's ability to accurately capture the flow around flapping wings during forward flight and estimate the corresponding aerodynamic loads. The flapping wing has a rectangular lifting surface with an aspect ratio, $A_R=8$, a flapping amplitude of $\delta =15°$ and a reduced frequency of $k=0.1$. As shown in Fig. 3, the deviation of our results from the previous study [51], [61] is limited to 4-6 % at angles of attack 0º and 4º. This demonstrates a strong agreement between the lift coefficients obtained from the current simulation using DVM and those obtained from the Euler solver [51] and UVLM [61].

### 4.3 Flow field generated by the flapping wing

The wake positions during a flapping cycle are depicted in Figure 4(a), with the blue and red lines indicating the vortices rotating in clockwise and counter clockwise directions, respectively. Spline interpolation [54] applied to the marker positions provide a more distinct representation of the wake shape (see Fig. 4a). A network of nodes located at specific positions (according to [68], the image plane is approximately $0.3c$ away from the trailing edge in the x-direction) can be utilized to compute the induced velocity (as shown in Fig. 4b). In previous experimental studies of flying animals' wakes [69]–[71], the velocity field has shown the counter-rotating wake regions behind the wingtips.

Fig. 4(c) and 4(d) show the side view of wake configurations after three cycles of symmetric and asymmetric flapping of rigid wings, respectively. The vortex rings induce velocity from their centre. The wake vortices propagate along the x-direction (refer to Fig. 4(c) and 4(d)). The wake's evolution was downward due to its self-induced velocity (see Fig. 4(c) and 4(d)).

It is clear that, in comparison to symmetric flapping (fig. 4c), the wake positions in asymmetric flapping (fig. 4d) are mostly located in the lower half (in the negative z-direction). Due to the predominance of downward fluid momentum and the resulting asymmetry in wake positions, the lift in asymmetric flapping is non-zero and positive. As a consequence of Newton's third law of motion, the wing experiences an upward lift force in the vertical direction. Moreover, this downward flow of vortices in the wake plays a significant role in generating horizontal force. This force is attributed to vortex impulse, which is linked to the change in momentum experienced by the wing as it sheds these vortices [72]. Vortex impulse is a key concept in vortex dynamics concerning the momentum carried by the trailing-edge vortices formed behind the wing. These vortices consist of discrete swirling air packets with specific circulation strength. When the wing imparts momentum to the vortices in its wake, there is an equal and opposite momentum change experienced by the wing, resulting in a horizontal force known as thrust. So, the asymmetric wing experiences higher thrust force due to the increased magnitude of vortex impulse.

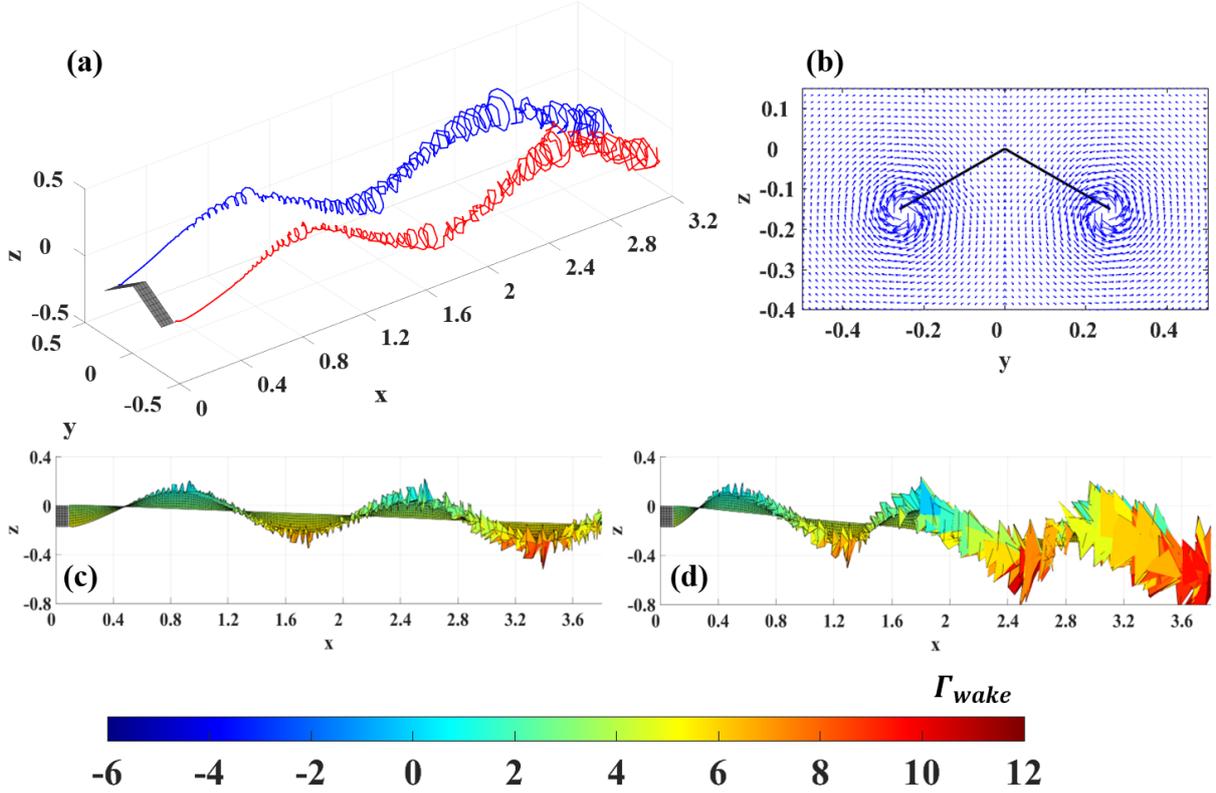

Fig.4. (a) Simulated wakes using cubic interpolation technique for symmetric ($A_S$=1) rigid wings(b) Induced velocity predicted on a uniform grid of nodes when viewed in the negative x-direction for symmetric ($A_S$=1) rigid wing (c) Side view of wake surfaces symmetric ($A_S$=1) flapping rigid wing (x-z plane). (d) Side view of wake surfaces for asymmetric ($A_S$=2) flapping rigid wing. Simulation results are obtained after 3 cycles of flapping motion at $\delta_1$ =30°; $\delta_0$=-30°; $k$=0.1; $A_R$=3.

## 4.1 Temporal variation of aerodynamic force generated during flapping motion

Fig. 5(a) and 5(b) show the temporal variation of $C_L = L/(1/2\ U_\infty^2)$, over 2 cycles of symmetric and asymmetric flapping, respectively, for the rigid, bending, twisting and BTC modes. The flapping stroke angle is 30° and the aspect ratio is chosen to be 3. The reduced frequency is set to 0.3, following [55]. The lift force is perpendicular to the free stream flow and parallel to gravity, but in the opposite direction.

According to previous studies [67], [73], [74]. it has been observed that in the flapping cycle, a major portion of the net lift is generated during the downstroke, whereas little lift force is produced during the upstroke. During the upstroke, birds may pull their wingtips in or increase sweep [75], reducing span, drag and power used among other things to improve propulsive efficiency. The magnitude of the lift is substantially changed when the twist is given in flapping motion (Fig. 5(a) and 5(b)) with the majority of the positive lift now occurring during the downstroke. This observation highlights the direct relationship between circulation (as depicted in Fig. 4(c) and 4(d)) and lift and thrust generation. Higher strength circulation leads to a more substantial pressure difference (Eq.10), resulting in increased lift and thrust forces, as demonstrated in Fig. 5(a) and 5(b). In case of asymmetric wing higher strength of circulation is observed during the downstroke which leads to increase the peak lift and thrust. When the wing is performing an upstroke motion, the value of lift decreases from $t/T_p=0$ to $t/T_p=0.25$ for symmetric flapping i.e., $A_S=1$ (for all modes). When the wing moves upward, the lift value begins to increase in a positive direction. The peak value of lift is achieved when the wing is in the middle of the downstroke ($0.75t/T_p$) as we have seen in earlier studies as well [51], [61]. In the case of symmetric flapping, lift of equal magnitude is generated in both the positive and negative directions, resulting in zero net lift.

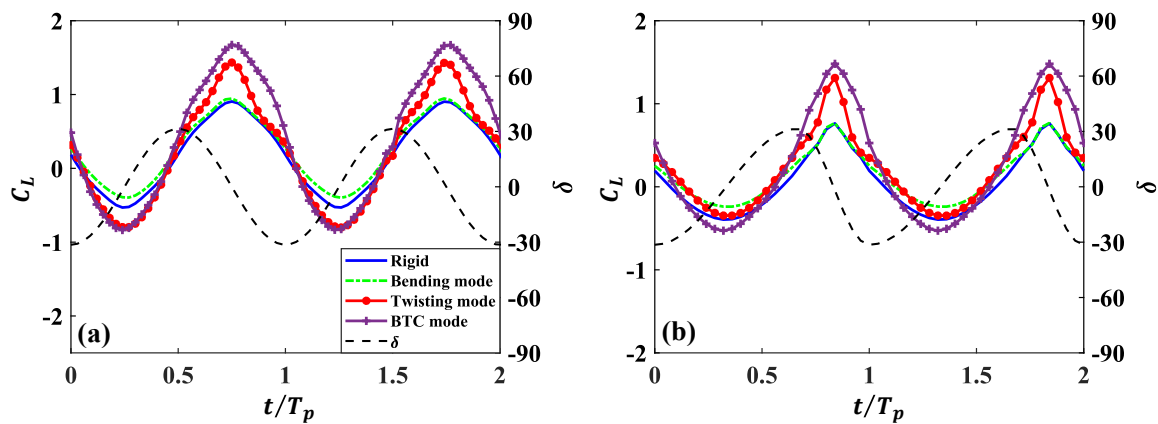

Fig. 5. (a) Lift coefficient curve over 2 cycles for symmetric (b) Lift coefficient curve over 2 cycles for asymmetric ($A_S$=2) ($A_S$=downstroke speed/upstroke speed=2) at $\delta_1$ =30°; $\delta_0$=-30°; k=0.3; $A_R$=3.

For asymmetric flapping ($A_S$=2), the downward stroke speed is twice the value of the upstroke speed. In this case, the overall lift coefficient is shifting upward. When the wing moves through a region of air that has a downward-directed vortex, the vortex can impact the airflow over the wing. Specifically, the downward flow of air in the vortex can increase the velocity of the air that flows over the upper surface of the wing. As the airspeed over the top of the wing increases, it can lead to a greater difference in pressure between the top and bottom surfaces of the wing. This pressure differential can result in a greater amount of lift being generated by the wing. In addition, for asymmetric cases, the rigid and shape-morphing wings' (twisted and BTC modes) peak amplitudes of positive lift are greater than those of negative lift. The ratios of maximal positive lift to maximal negative lift are approximately 1.4 and 2.5 for the twisting mode of symmetric and asymmetric flapping, respectively. The lift curve shows no discernible difference for rigid and bending modes in both the symmetric and asymmetric flapping cases. In the symmetric case, the peak lift in twisting and BTC mode is approximately 1.51 and 1.77 greater than that of a rigid symmetric, while in the asymmetric case, the peak lift in twisting and BTC mode is approximately 1.72 and 1.94 higher than a rigid asymmetric. Overall, the lift curve exhibits similar trends for rigid and bending modes in symmetric and asymmetric flapping, but demonstrates higher positive lift for twisting and BTC modes.

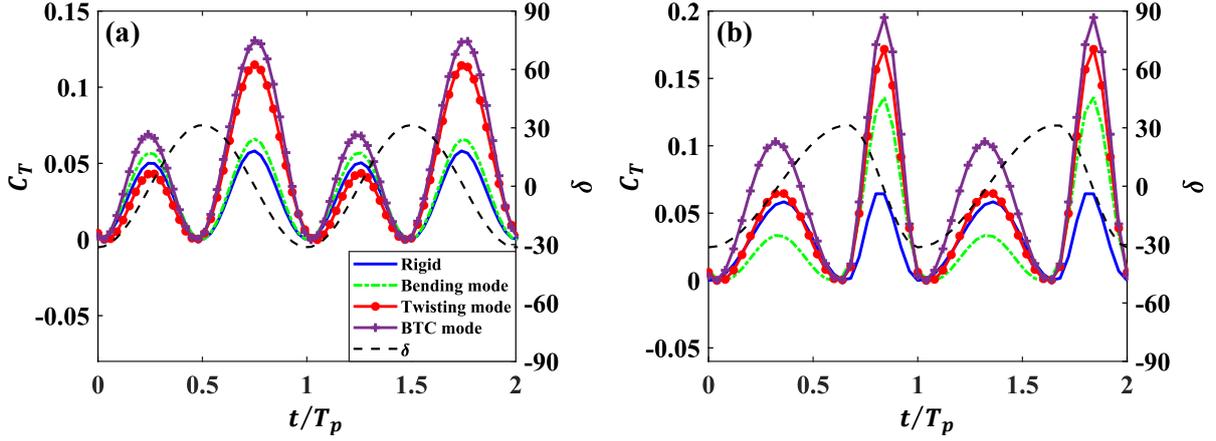

Fig. 6. (a) Thrust coefficient curve over 2 cycles for symmetric (b) Thrust coefficient curve over 2 cycles for asymmetric ($A_S$=2) ($A_S$=downstroke speed/upstroke speed=2) at $\delta_1$ =30°; $\delta_0$=-30°; $k$=0.3; $A_R$=3.

Subsequently, Figures 6a and 6b depict the temporal evolution of thrust coefficient ($C_T = T/(1/2\ U_\infty^2)$). It should be observed that the thrust force peaked during both the upstroke and the downstroke. The majority of thrust production occurs during the downstroke. During the upstroke, some drag (negative thrust) is produced. In both cases of Figure 5a and b, no discernible enhancement of lift was observed in the case of bending modes compared to the rigid case. However, as shown in Figure 6b, it is evident that the thrust during bending mode is significantly higher than in the rigid case. In the symmetric case, the peak thrust in bending, twisting and BTC mode is approximately 1.137, 1.72 and 1.96 greater than that of a rigid symmetric, while in the asymmetric case, the peak thrust in bending, twisting and BTC mode is approximately 2.12, 2.67 and 3.05 higher than a rigid asymmetric. For both symmetric and asymmetric flapping, the use of BTC appears to be the most effective method for generating thrust. During the symmetric bending mode, the peak is 0.0660, which is 1.15 times higher than the peak during the upstroke. The asymmetric bending mode exhibits a peak of 0.1359 during the downstroke, which is 4.06 times greater than the peak observed during the upstroke. In the case of the symmetric twisting mode, the peak during the downstroke is 0.1147, which is 2.67 times greater than the peak during the upstroke. For an asymmetric twisting mode with $A_S = 2$, the peak in the downstroke is approximately 0.171, which is 2.61 times higher than the peak in the upstroke. Compared to a rigid wing, the twisting and bending modes show an overall increase of 57.9% and 22.9%, respectively, in the symmetric case. In the case of asymmetry, the twisting and bending modes increase by a factor of 2.13 and 1.53, respectively, when compared to a rigid wing. When a wing experiences a BTC mode, it generates a peak thrust 2.23 and 3.04 times greater than that of a rigid wing in a symmetric and asymmetric configuration, respectively.

The higher flapping speed during the downstroke is believed to be the reason for the distinct force peaks. Thus, the thrust can be modulated by both wing flapping and wing deformation. Additionally, the peak of the downstroke $C_T$ was observed in the middle of the downward stroke. While the thrust is significantly reduced during the downstroke, these forces also impose a significant drag penalty during the upstroke. When a wing is twisted, the angle of attack varies across the span of the wing. This means that various sections of the wing encounter varying angles of attack, which can lead to fluctuations in the lift and drag forces exerted on the wing. The overall impact of these variations is that the wing is capable of producing greater lift and reduced drag, ultimately leading to increased thrust.

## 4.2 Pressure distribution over the wing

Figures 7(a)-(d) show the pressure distribution on the wing surface at various positions during the downstroke and upstroke motions of symmetric flapping for the four modes: rigid wings, bending, twisting, and BTC. The pressure is coefficient of an individual panel ($i,j$) is expressed as, $Cp_{ij} = p_{ij}/(1/2 \; \rho U_\infty^2)$, where the pressure over the body is assumed to be gauge pressure. Due to the force exerted on the fluid by the wings, an equal and opposite force is exerted on the body in accordance with Newton's third law of motion. Hence, the velocity of the jets is an indication of thrust production. This section essentially provides justification for the temporal variation of lift and thrust. It explains how pressure changes during the downstroke and upstroke motions, resulting in alterations in lift (Fig.5a) and thrust (Fig.6a) production. In the rigid case, there is a positive pressure during the downstroke and a negative pressure during the upstroke, which mirrors the same pattern observed in the temporal variation of lift and thrust. The lift (Fig.5a) and thrust (Fig.6a) curves show their peak magnitudes at the mid position of the wing, highlighting a similar trend in pressure magnitude. In the case of bending, the pressure magnitude increases slightly compared to the rigid case; however, the pressure magnitude during the downstroke and upstroke remains quite similar. In the case of twisting and BTC, there's an overall increase in pressure. Additionally, the positive pressure magnitude during the downstroke surpasses that during the upstroke, resulting in an overall higher pressure throughout the entire cycle. The pressure peak, facilitating the attachment of airflow around the leading edge of the wing, can lead to an escalation in the adverse pressure gradient and serve as a sign of an imminent flow [53], [76].

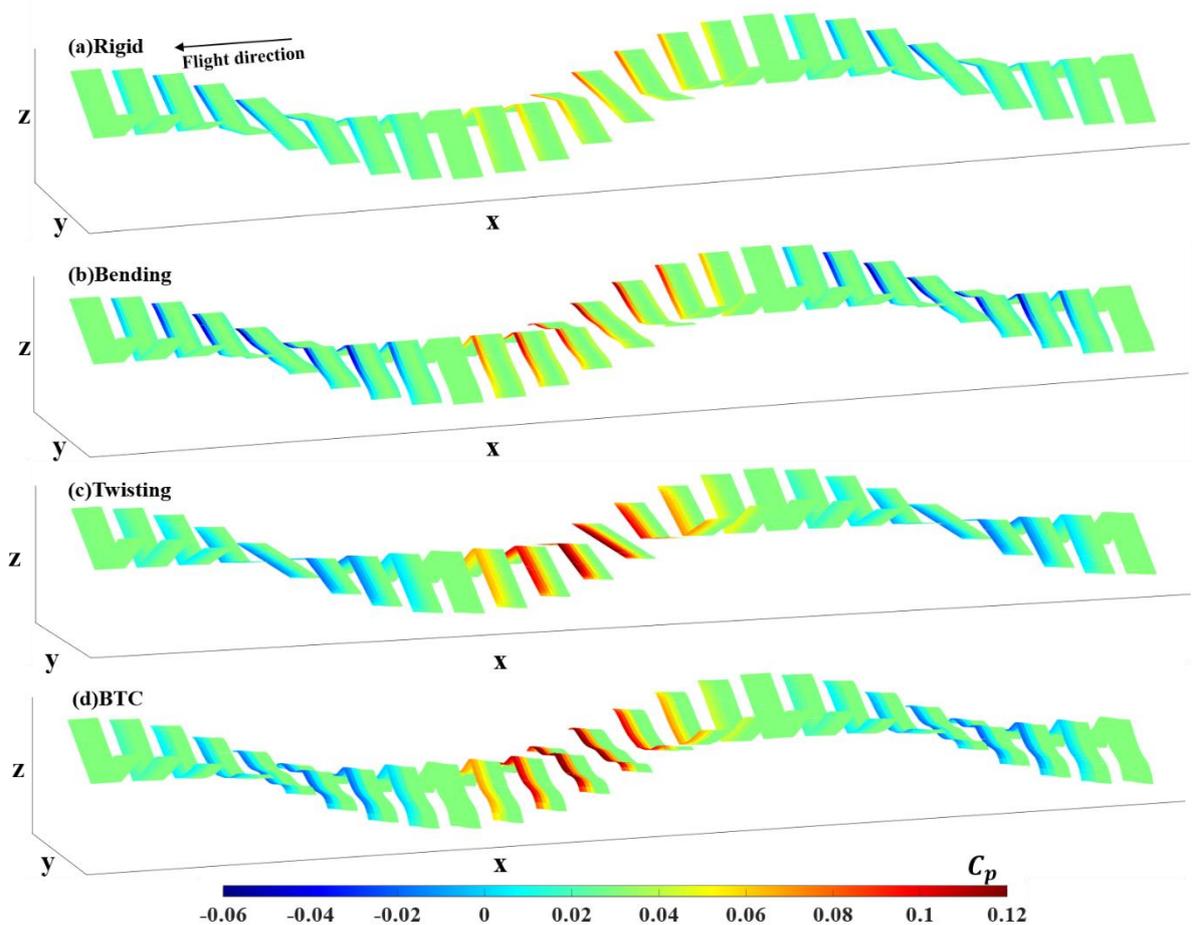

Fig.7. Pressure coefficient distribution over 1.5 cycles for the symmetric flapping wing at $\delta_1 =30°$; $\delta_0=-30°$; $k=0.3$; $A_R=3$; $A_S=1$ (a) Rigid (b) Bending (c)Twisting and (d)BTC.

## 4.3 Influence of reduced frequency on the flapping wing

The reduced frequency ($k$) significantly affects the unsteady aerodynamics of flapping wings. The reduced frequency determines how quickly the wing is flapping in relation to the speed of the surrounding fluid. At higher reduced frequencies, the unsteady term $\left(\frac{\partial}{\partial t}\Gamma_{ij}\right)$ in equation (11) plays a significant role in lift and thrust calculations [58]. Furthermore, since the majority of the wake is concentrated near the wing, local fluctuations in the wake terms that govern the no-penetration condition become more significant [61]. Figures 8 a-d illustrate the trailing-edge vortices circulation strength of the wakes obtained for the rigid, bending, twisting and BTC wing respectively. These trailing edge vortices are caused by the high-pressure air at the bottom of the wing moving towards the low-pressure air above the wing, which creates a rolling motion of air particles. At each time step, vortices are shed from the trailing edge of the wings and convected in a negative x-direction. Trailing-edge vortices create an induced velocity field in the surrounding flow. The induced velocity field represents the velocity that is generated due to the presence of the vortices. This induced velocity affects the flow field $(u, v, w)_w$, leading to lift enhancement (refer eq 10). Comparison of the wake trajectory of all modes shows that the wake vortices is organized for the rigid wing in comparison to the deforming wing. In the case of the twisted and BTC modes, the wake trajectory exhibits chaotic behavior, characterized by irregular and unpredictable flow patterns. This chaotic wake results from the complex interactions between the wing's motion and the vortices shed from its trailing edge. The twisting motion introduces non-uniform vorticity distribution along the wing's span, leading to a more disordered wake flow. Moreover, in these modes, the magnitude of circulation is higher compared to other configurations. Circulation represents the strength of the swirling motion of fluid particles around a closed curve. In the twisted and BTC modes, the wing's motion induces higher circulation in the trailing-edge vortices, resulting in more intense and stronger rotational characteristics. As per the Kelvin's circulation theorem [58], the combined circulation around a flapping wing and its wake remains constant over time $\left(\frac{D\Gamma_{wing}}{Dt} + \frac{D\Gamma_{wake}}{Dt} = 0\right)$. Therefore, if the magnitude of the wake's circulation increases, the magnitude of the wing's circulation also increases correspondingly. This conservation principle ensures that changes in the wake's circulation directly influence the circulation of the wing itself. The relationship between wake and wing circulation is crucial in understanding the aerodynamic performance of the deformable wing in these modes. An increase in wake circulation increases the pressure distribution and flow patterns around the wing, influencing lift and thrust generation. Furthermore, the wing's lower surface exhibits a high-pressure zone, while the upper surface demonstrates a low-pressure zone. As the wing twists, the high-pressure zone may abruptly reconnect with the surface of the wing, resulting in a sudden release of energy into the flow and intensifying the vortex. The effect enhances the strength of the circulation in a wing that is twisted. This high strength circulation imparting substantial effects on the induced velocity field. This phenomenon, in turn, leads to a notable increase in both lift and thrust force generated by the wing [61], [65]. Therefore, we can conclude that applying a twist is the most effective method for generating lift when compared to other techniques.

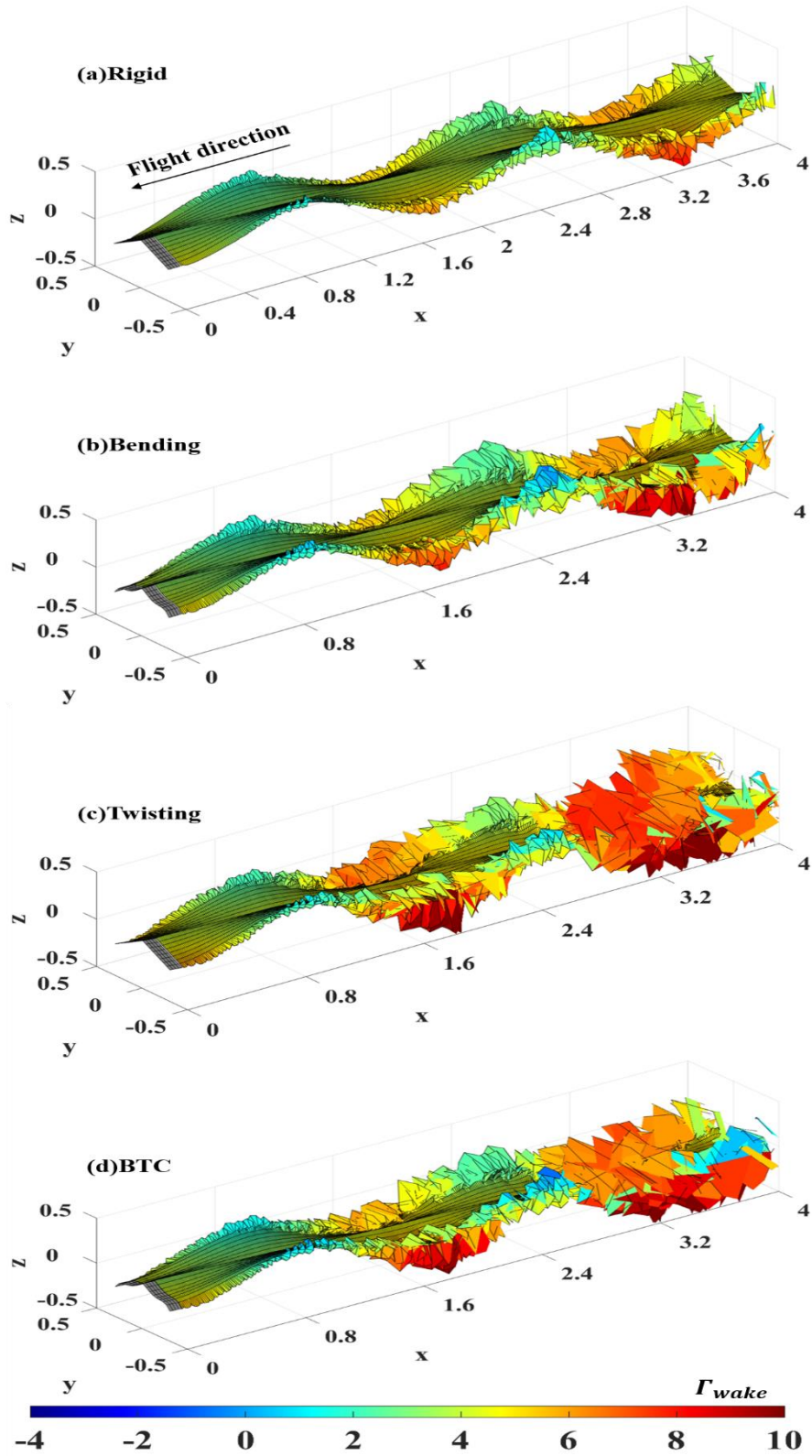

Fig.8. Isometric view of wake surfaces after 3 cycles for flapping wing at $\delta_1 = 30°$; $\delta_0 = -30°$; $k=0.1$; $A_R=3$; $A_S=1$. (a) Rigid (b) Bending mode (c) Twisting and (d) BTC mode.

To gain a clearer understanding of the wake circulation near the wing surface, we compare the rigid, bending, twisting, and BTC modes in Figure SI -I. By examining the rigid (Figure

SI-1a) and bending (Figure SI-1b) modes, we can observe a reduced presence of circulation near the wingtip, leading to decreased drag. The twisted (fig.SI-1c) and BTC wing (fig.SI-1d) help to maintain the circulation of the higher strength vortex near to the trailing edge of the wing (Fig. SI-I). When a wing is twisted, the angle of attack varies along its length, resulting in varying lift distributions at different locations [77]. When the angle of attack increases towards the wingtip, there is a greater pressure differential between the upper and lower surfaces of the wing. This creates a more concentrated vortex and stronger circulation towards the wingtip due to a stronger pressure gradient. As a result, the twisted wing design results in a higher lift. As shown in Figure 5, the twisted and BTC wings exhibit a higher peak lift magnitude compared to the rigid and bending wings.

Figure 9 illustrates the wakes for rigid, bending, twisting and BTC cases at a higher frequency ($k$=0.6), allowing for clear visualization of the circulation strength of the wakes. When the reduced frequency value ($k$) is lower, such as $k = 0.1$, the flapping dynamics resemble those observed in bird flight. However, at a higher reduced frequency value, like $k = 0.6$, the motion becomes more similar to hovering behavior [78], [79]. The vortex rings that form in the wake after a flapping cycle are located quite close to the wing. The aerodynamic behavior of a flapping wing is significantly influenced by the interactions between the wing and the wake. The vortex rings in the wake behind the wing with a higher reduced flapping frequency are higher in magnitude (fig.9), indicating an increase in lift compared to the wing with a lower reduced frequency ($k$=0.1). At higher reduced frequencies, the aerodynamic performance of a flapping wing tends to improve because the wing's flapping motion generates an increased pressure level (refer SI-II). The higher dynamic pressure enhances lift and thrust production, which enables greater maneuverability. Moreover, higher reduced frequency pertains to a more rapid rate of vortex shedding from a trailing edge (fig.9), typically occurring in unsteady flow conditions. This increased frequency leads to more frequent formation, amplification, and shedding of vortices. Consequently, the wake exhibits energetic flow patterns due to the interaction of these vortices. The enhanced shedding and energy transfer processes intensify the swirling motion and circulation. As a result, higher frequency facilitates the augmentation of lift and thrust generation, making it advantageous in specific flight regimes for improved aerodynamic performance. Furthermore, the vortex located above the deformed wings (bent, twisted and BTC) is significantly weaker than the rigid case when $k$=0.6 (Fig. 9), which is indicative of a large positive lift. For rigid wings, the negative vortex located above the wing has higher magnitude within the wake that occurs during both strokes (fig 9b), resulting in a negative lift penalty for the wing. In the case of a wing undergoing deformation, the positive vortex is deliberately positioned beneath the wing (as shown in Figures 9d, 9f, and 9h), resulting in a consistently high average lift value.

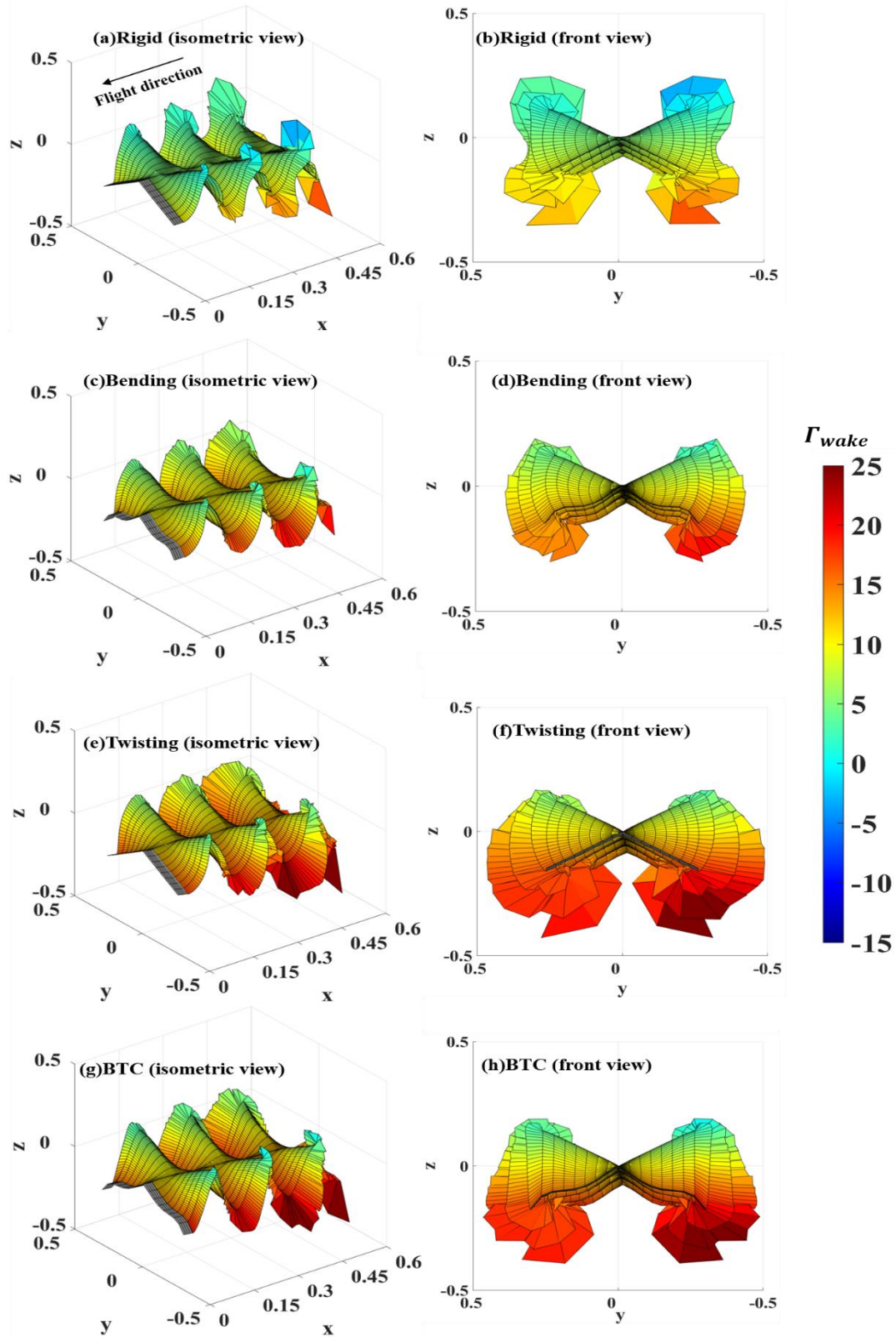

Fig. 9. Shows the isometric and front views of wake surfaces after 3 cycles for a flapping wing with various configurations. The wing parameters are as follows: $\delta_1 = 30°$; $\delta_0 = -30°$; $k=0.6$; $A_R=3$; $A_S=1$. The different configurations include rigid, bending, twisting, and BTC.

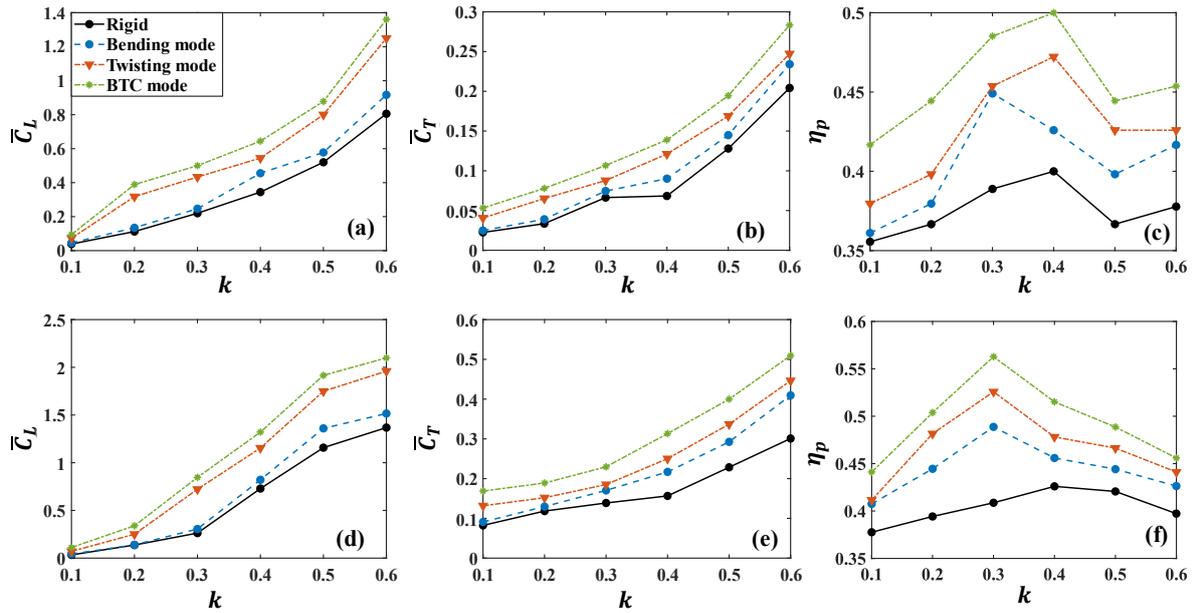

Fig. 10. Influence of reduced frequency on $\bar{C}_L$, $\bar{C}_T$ and $\eta_p$. results are obtained after 10 cycles of flapping motion at $\delta_1 = 30°$; $\delta_0 = -30°$; $A_R = 3$; (a)-(c) represents symmetric flapping; (d)-(f) represents asymmetric flapping ($A_S = 2$).

Fig.10 shows the effect of reduced frequency on time average lift ($\bar{C}_L$), time average thrust ($\bar{C}_T$) and propulsive efficiency ($\eta_p$) for the four different wing configurations. Researchers can enhance their comprehension of the aerodynamic performance of flapping wings through the application of reduced frequency analysis [80], [81]. Here, we are testing the performance of the wing at six different reduced frequencies. As the reduced frequency ($k$) increases, the aerodynamic forces (lift and thrust) at each time step are significantly influenced by the wing shape in the previous time steps. This could be attributed to the influence of the third term ($\frac{\partial}{\partial t}\Gamma_{i,j}$) from eqn.10, which becomes prominent at high reduced frequency, and its significant contribution in determining the aerodynamic force [58]. As the reduced frequency increases, the velocity of the wing tip also increases. The range of the reduced frequency is from 0.1 to 0.6. It is readily evident form figures 10 a and b, $\bar{C}_L$ increases with increase in $k$. First, we compare the performance of four types of wings (rigid, bending, twisting and BTC) in terms of lift coefficient at a higher reduced frequency ($k = 0.6$) for both symmetric (fig. 10a) and asymmetric case (fig. 10d). In the case of symmetric and asymmetric bending modes, the resulting lift coefficients are 0.916 and 1.51, respectively. This represents a 13.78% and 11% increase compared to the rigid symmetric (0.805) and asymmetric (1.36) modes at $k=0.6$. In the case of both symmetric and asymmetric twisting mode, the resulting lift coefficient is substantially higher i.e., 55.25% and 43.3% more than in the rigid symmetric and asymmetric modes. The lift coefficient increases significantly in both symmetric and asymmetric BTC modes, with an increase of 68.75% ($\bar{C}_L=1.35$) and 53.57% ($\bar{C}_L=2.15$) compared to the lift coefficient in the rigid symmetric and asymmetric modes. This analysis shows that the BTC mode can significantly enhance the lift performance of flapping wings. As the reduced frequency increases, the thrust force increases. In the case of a twisted wing, the thrust force is significantly greater than that experienced by both rigid and bending modes. Upon scrutinizing the wake topology for all four cases (Figures 8a-d), it is evident that the magnitude of circulation in a BTC wing is significantly greater. Fig. 10(c) and 10(f) demonstrate the impact of reduced frequency on propulsive efficiency for symmetric and asymmetric flapping, respectively. The propulsive efficiency of a wing is influenced by various factors, such as the wing's geometry, flow conditions, and operating conditions. One of the key factors that affects the propulsive efficiency of a wing is the reduced frequency ($k$) of the wing. Eqn. 13 defines the mathematical expression for propulsive efficiency. This has been observed that the propulsive efficiency of a wing increases significantly with the reduced frequency up to a maximum value in the moderate range of reduced frequencies, typically between 0.3 and 0.4. After reaching its maximum value, propulsive efficiency begins to monotonically decrease as the reduced frequency increases further.

### 4.4 Influence of aspect ratio ($A_R$)

Fig. 11 highlights the influence of the aspect ratio ($A_R$) on propulsive efficiency. Keeping the area of the wing constant, the $A_R$ has been changed from 1 to 6 to investigate the impact of wing shape on propulsive efficiency. For both symmetric and asymmetric flapping, all the four modes exhibited same variation of $C_p$ versus $A_R$. The propulsive efficiency increases continuously until reaching $A_R$~3-4 and subsequently decreases with a further increase in $A_R$. In a previous study on flapping foils [64], it has also been reported that when the $A_R$ exceeds a certain limit ($A_R$=2.55), the vortex starts to detach from the wingtip resulting in a drop in the efficiency. These findings demonstrate wings with high aspect-ratio ($A_R$~3-4) can be selected for optimal aerodynamic performance. The angular velocity at the wing tip are higher for wings with high $A_R$. Higher aspect ratio wings experiencing greater deformation can contribute to improved aerodynamic performance.

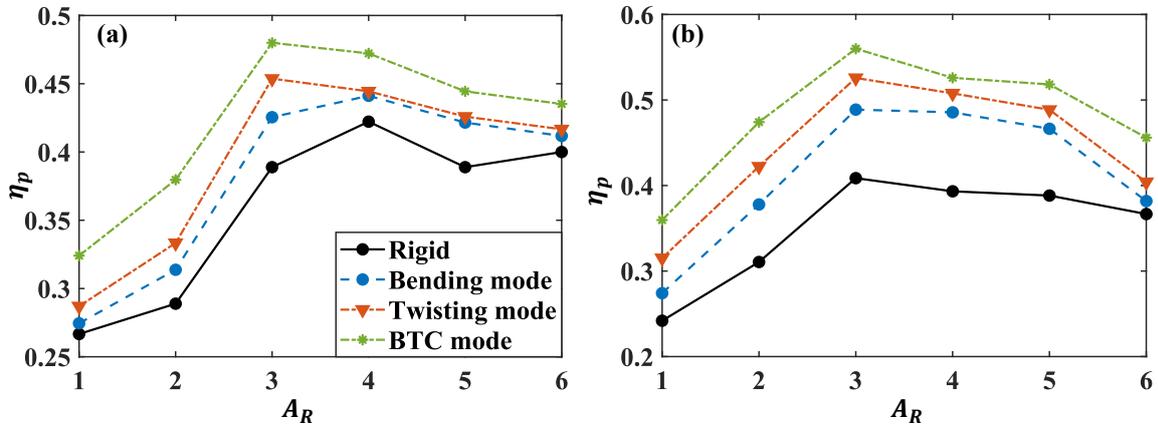

Fig. 11. Influence of aspect ratio on propulsive efficiency (a)symmetric (b) asymmetric ($A_S$=2). results are obtained after 10 cycles of flapping motion at $\delta_1$ =30°; $\delta_0$=-30°; $k$=0.3.

### 4.5 Influence of asymmetric ratio ($A_S$)

As evident from figure 12, the $A_S$ has a significant impact on the propulsive efficiency ($\eta_p$) for all four cases. In all cases, propulsive efficiency increases with the asymmetric ratio and reaches a maximum near the range of $A_S$=2 to 3, after which it monotonically decreases. These findings demonstrate that altering the asymmetric ratio has an impact on propulsive efficiency. In flapping wings, the downstroke is generally considered the power stroke. When the speed of the downstroke is higher, the wing can generate a greater downward force on the air, resulting in a larger pressure differential and a greater amount of lift (refer fig.SI-VI (a)). Additionally, a higher downstroke speed can increase the thrust force (refer fig.SI-VI (b)). This is because the motion of the wing through the air generates vortices that trail behind it.

A higher downstroke speed can result in larger and stronger circulation vortices, which can increase the wing's thrust force. The propulsive efficiency initially increases as the asymmetric ratio increases up to 2, but then it decreases monotonically. The increase in pressure resulting from the increase in asymmetric ratio is responsible for the decrease in propulsive efficiency. Consequently, this leads to an overall reduction in propulsive efficiency.

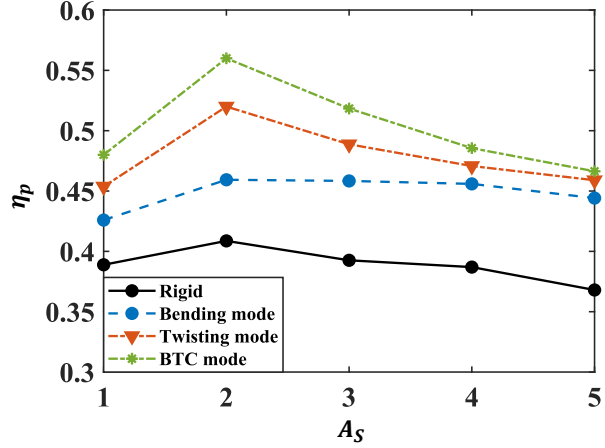

Fig.12. Influence of asymmetric ratio on propulsive efficiency at $k$=0.3. Results are obtained after 10 cycles of flapping motion at $\delta_1$ =30°; $\delta_0$=-30°; $A_R$=3.

## 5. Conclusion

The present study involves numerical simulation of the 3D flapping motion of a rectangular wing using the DVM scheme. We validated our DVM code by comparing it to a previous study on a rectangular wing in forward flight with a flapping motion. we have conducted a comparative assessment of the aerodynamic performance across four distinct configurations: rigid, bent, twisted, and BTC (bend-twist-coupled) wings. The current study is centered on the analysis of lift ($C_L$) thrust ($C_T$), and propulsive efficiency ($\eta_p$), along with the associated vortex dynamics, of a three-dimensional flapping wing. We conducted a parametric study to investigate the dependencies of $C_L$, $C_T$, and $\eta_p$ on various kinematic parameters, including aspect ratio, asymmetric ratio, and reduced frequency. Initially, we compared the performance of symmetric and asymmetric wing motion, wherein we observed that in the case of asymmetric wing motion, the time-averaged aerodynamic forces exhibited a notable increase. Furthermore, we conducted a comparative analysis among the bending, twisting, and BTC wings in comparison to the rigid wing. In the case of the bending wing, there was a remarkable increase in the overall lift force, although thrust improvement was relatively modest. On the other hand, both the twisting and BTC wings demonstrated more effective generation of both lift and thrust forces. Subsequently, we scrutinized the pressure distribution across the wing surfaces for all four cases and observed that the highest pressure levels were attained by the BTC wing. Consequently, this yielded the highest aerodynamic forces for the BTC wing. Furthermore, we compared the performance of all four wing configurations across different reduced frequencies ($k$) and asymmetric ratios ($A_S$). We noted that both lift and thrust increased with increasing $k$, and propulsive efficiency displayed an initial increase then gradually decreasing, with maximum efficiency observed in the range of $k$ = 0.3 to 0.4. Furthermore, the aspect ratio emerged as a pivotal factor influencing wing performance. Higher aspect ratios ($A_R$) correlated with elevated tip velocities, consequently augmenting the aerodynamic forces exerted on the wing. Lastly, we plotted a graph of propulsive efficiency vs asymmetric ratio (AS). In this analysis, we observed a interesting pattern where propulsive efficiency reached its peak at AS = 2, followed by a subsequent reduction in efficiency as the asymmetric ratio was increased. In summary, this numerical study serves as an outstanding showcase of employing discrete vortex methods to simulate three-dimensional unsteady flapping dynamics. It is our conviction that this study can establish a valuable framework for the development of efficient configurations in the quest for optimizing the design of flapping micro air vehicles (MAVs).

**Acknowledgments:**
We would like to express our gratitude to the Aeronautics Research & Development Board (ARDB) under the Defence Research & Development Organisation (DRDO), Government of India, is acknowledged by the authors for providing support through Research Grant No. ARDB/01/1031935/M/I.

**Data Availability statement:**
The data that support the findings of this study are available from the corresponding author upon reasonable request.

Aerosp. Sci. Technol., vol. 80, pp. 354–367, 2018, doi: 10.1016/j.ast.2018.07.017.

# Supplementary section

# Aerodynamic performance and flow mechanism of 3D flapping wing using discrete vortex method

**Rahul Kumar** [a], **Srikant S. Padhee** [a] and **Devranjan Samanta** [a,*,1]

[a] *Department of Mechanical Engineering, Indian Institute of Technology Ropar Rupnagar-140001, Punjab, India*

*Corresponding author:

[1]E-mail: devranjan.samanta@iitrpr.ac.in

[1]Tel: +91-1881-24-2109

## 1. Wake surface near the symmetric wing (*k*=0.1)

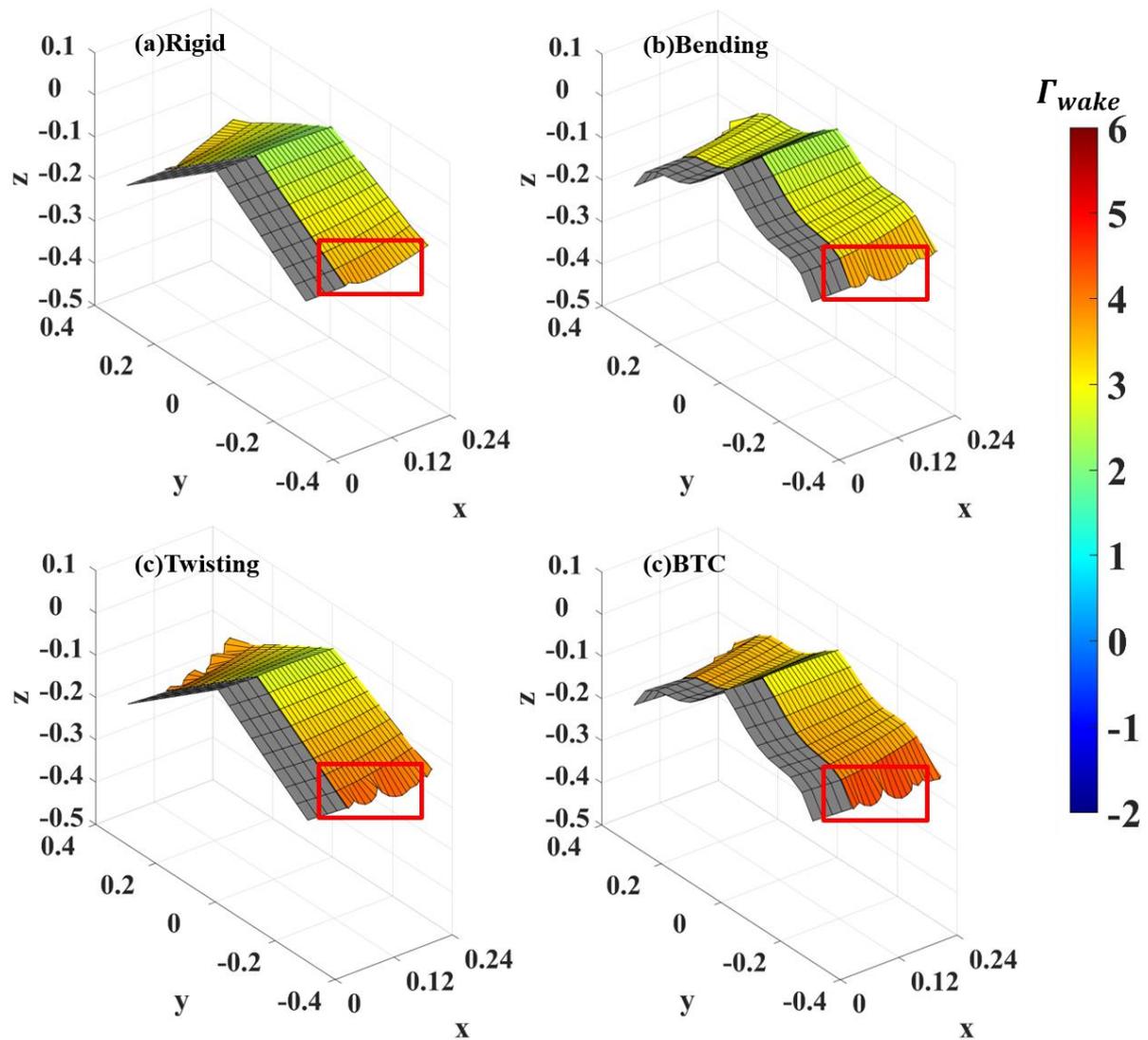

Fig. SI-I. - Wake surface near the wing is attached to the surface of wing (inside the red box) for flapping wing at $\delta_1$ =30°; $\delta_0$=-30°; *k*=0.1 after 3 cycle*s*. (a)Rigid (b) Bending mode (c) Twisting and (d) BTC mode.

## 2. Pressure distribution over the wing (*k*=0.6)

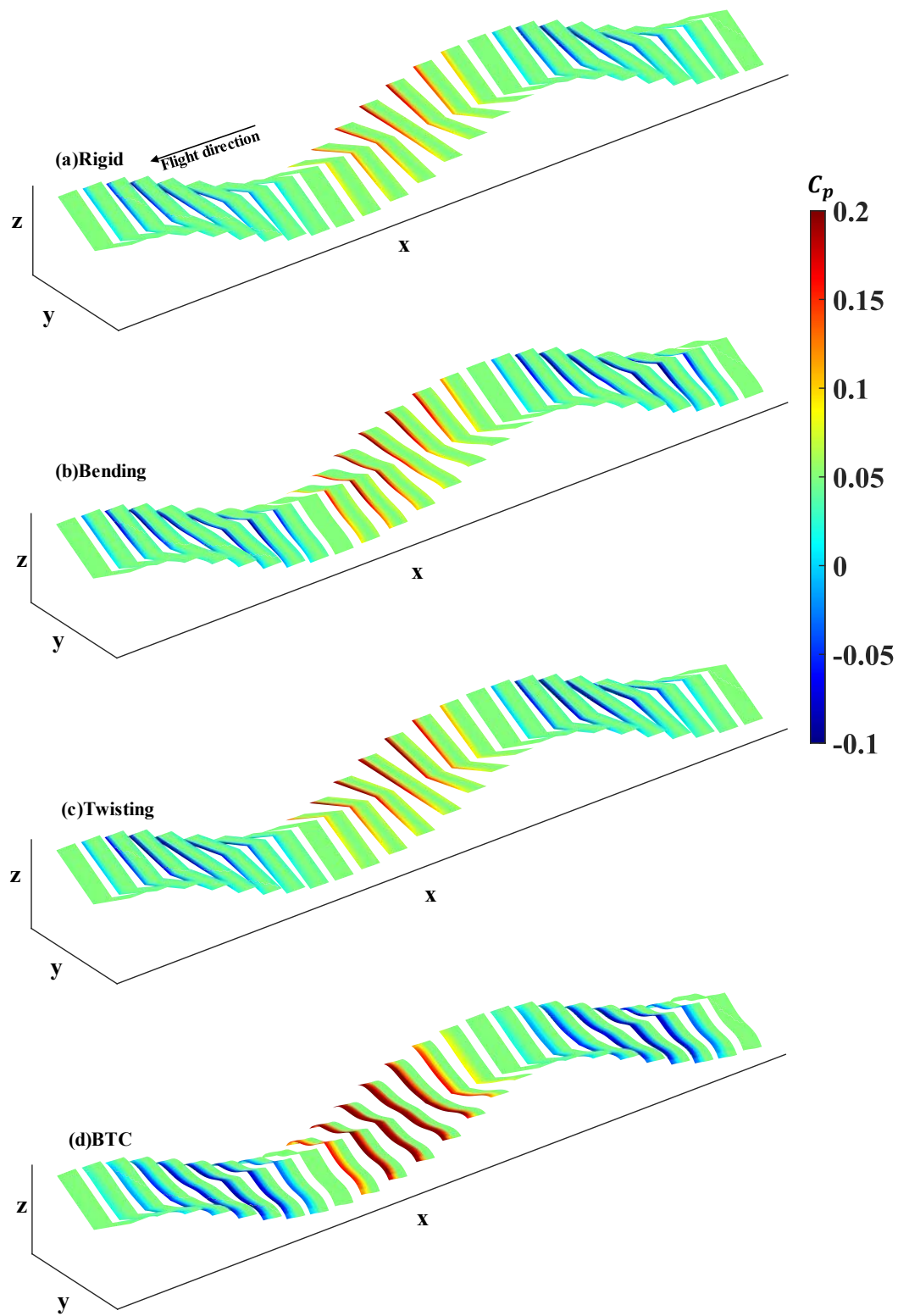

Fig. SI-II. Pressure coefficient distribution over 1.5 cycles for the symmetric flapping wing at $\delta_1 = 30°$; $\delta_0 = -30°$; $k=0.6$; (a) Rigid (b) Bending (c)Twisting and (d)BTC.

## 3. Wake surface for asymmetric wing ($k=0.1$)

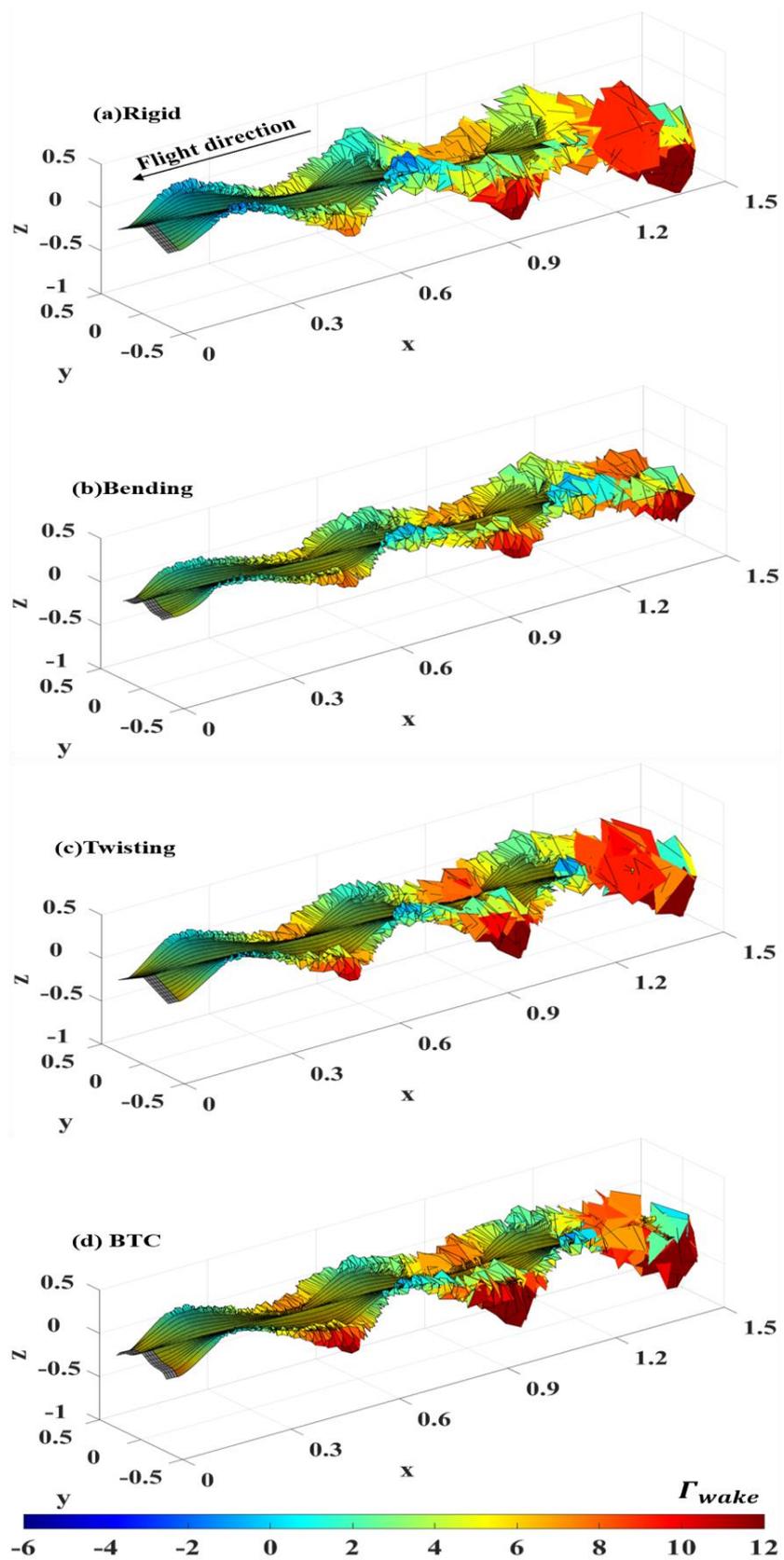

Fig. SI-III. Isometric view of wake surfaces after *3 cycles* for flapping asymmetric wing at $\delta_1 = 30°$; $\delta_0 = -30°$; $k=0.1$; $A_R=3$; $A_S=2$. (a)Rigid (b) Bending mode (c) Twisting and (d) BTC mode.

## 4. Wake surface for asymmetric wing (*k=0.6*)

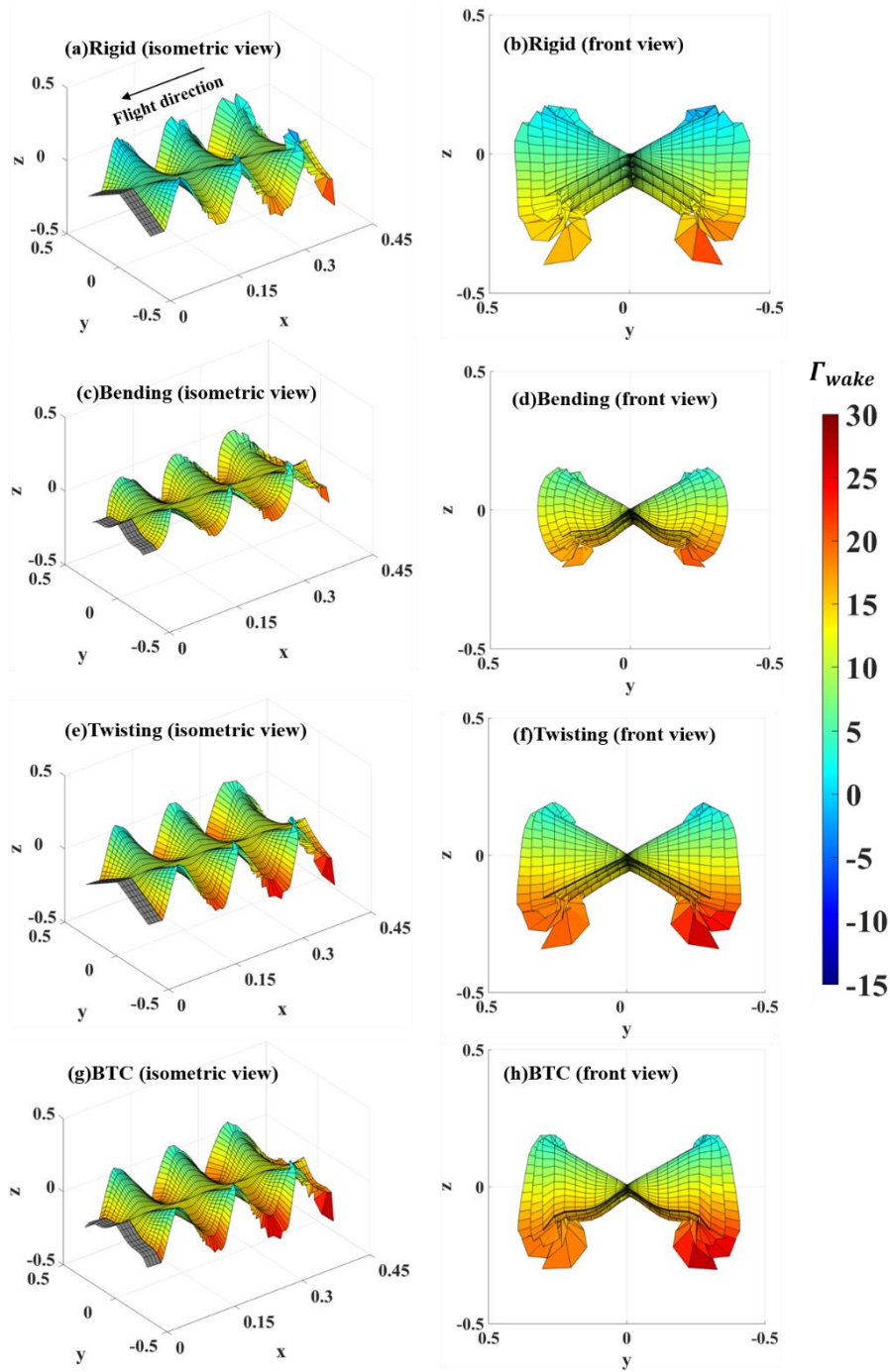

Fig. SI-IV. shows the isometric and front views of wake surfaces after 3 cycles for a flapping wing with various configurations. The wing parameters are as follows: $\delta_1 = 30°$; $\delta_0 = -30°$; $k=0.6$; $A_R=3$; $A_S=1$. The different configurations include rigid, bending, twisting, and BTC.

## 5. Influence of aspect ratio ($A_R$)

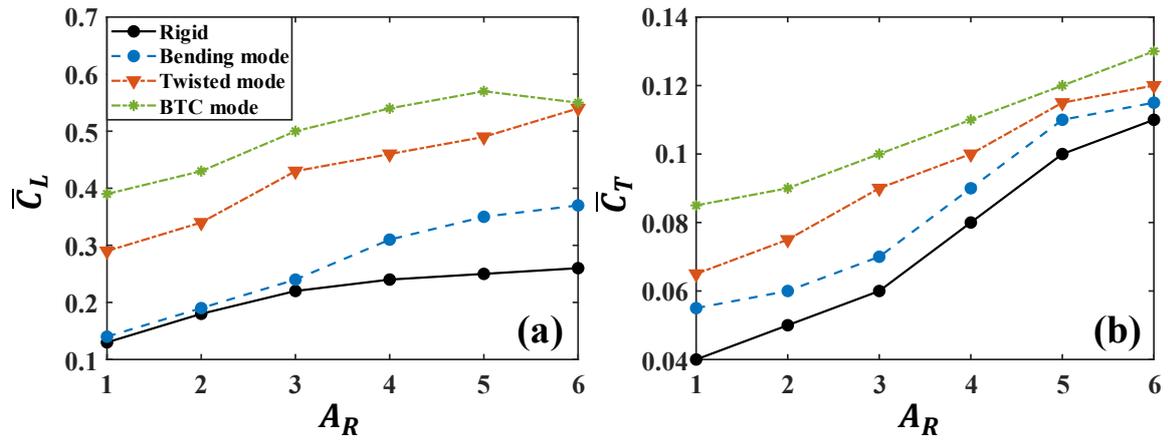

Fig. SI-V. Influence of aspect ratio on symmetric wing (a) $\bar{c}_L$ and (b) $\bar{c}_T$ results are obtained after 10 cycles of flapping motion at $\delta_1 = 30°$; $\delta_0 = -30°$; $k=0.3$; $A_R=3$.

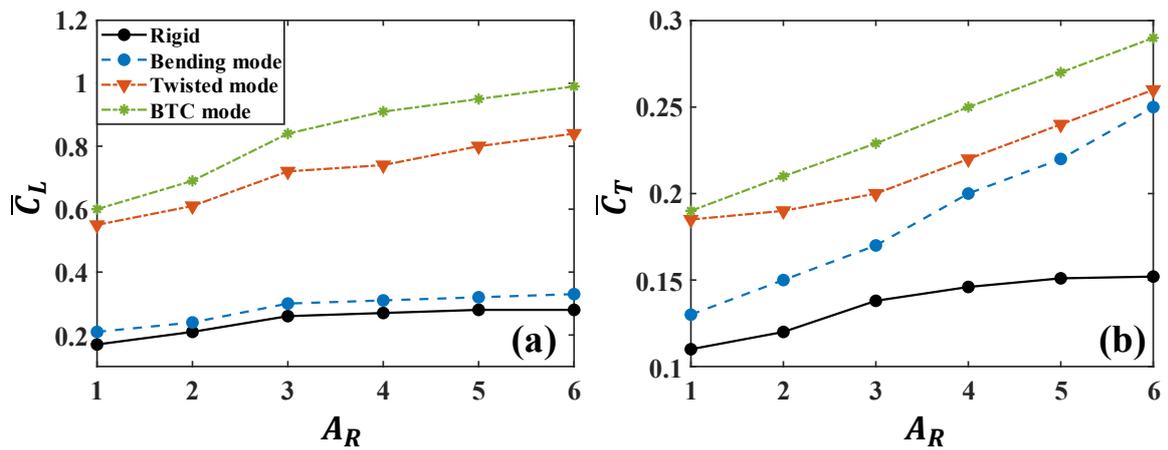

Fig. SI-VI. Influence of aspect ratio on asymmetric wing (a) $\bar{c}_L$ and (b) $\bar{c}_T$ results are obtained after 10 cycles of flapping motion at $\delta_1 = 30°$; $\delta_0 = -30°$; $k=0.3$; $A_R=3$.

## 6. Influence of asymmetric ratio ($A_S$)

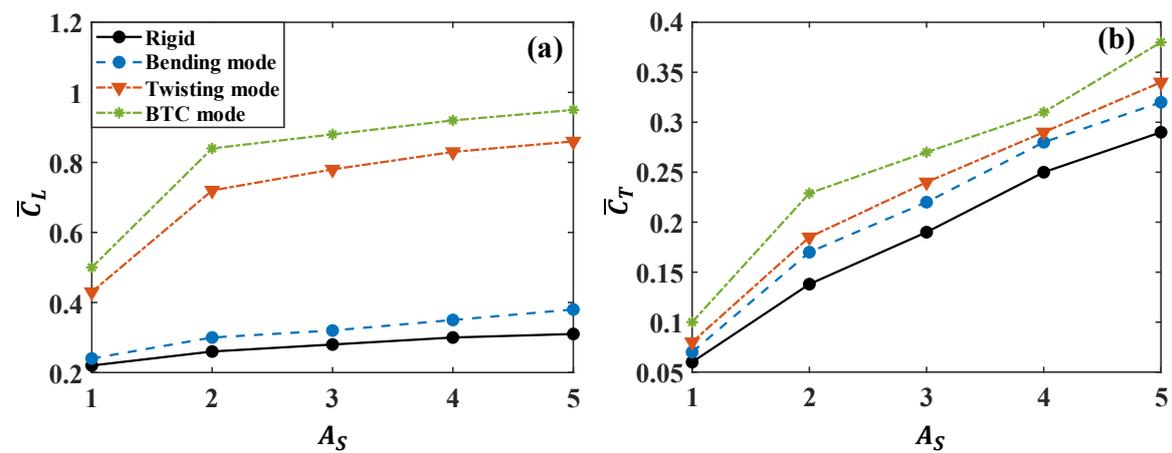

Fig. SI-VII. Influence of aspect ratio on asymmetric wing (a) $\bar{c}_L$ and (b) $\bar{c}_T$. results are obtained after 10 cycles of flapping motion at $\delta_1 = 30°$; $\delta_0 = -30°$; $k = 0.3$; $A_R = 3$.